\documentclass[twocolumn]{aastex62}

\usepackage{IEEEtrantools} 
\usepackage{amsmath}

\newcolumntype{L}[1]{>{\raggedright\arraybackslash}p{#1}}
\newcolumntype{C}[1]{>{\centering\arraybackslash}p{#1}}
\newcolumntype{R}[1]{>{\raggedleft\arraybackslash}p{#1}}

\newcommand{\msun}{\ensuremath{M_{\odot}}}
\newcommand{\lsun}{\ensuremath{L_{\odot}}}
\newcommand{\rsun}{\ensuremath{R_{\odot}}}
\newcommand{\rhosun}{\ensuremath{\rho_{\odot}}}
\newcommand{\numax}{\ensuremath{\nu_{\textrm{max}}}}
\newcommand{\dnu}{\ensuremath{\Delta\nu}}

\newcommand{\teff}{\ensuremath{T_{\textrm{eff}}\:}}
\newcommand{\logg}{\mbox{$\log g$}}
\newcommand{\muHz}{\mbox{$\mu$Hz}}
\newcommand{\feh}{\mbox{$\rm{[Fe/H]}$}}
\newcommand{\vsini}{\mbox{$v\sin i$}}

\newcommand{\rhostar}{\mbox{$\rho_{\star}$}}
\newcommand{\rstar}{\mbox{$R_{\star}$}}
\newcommand{\lstar}{\mbox{$L_{\star}$}}
\newcommand{\mstar}{\mbox{$M_{\star}$}}
\newcommand{\massp}{\mbox{$M_{\rm p}$}}
\newcommand{\radp}{\mbox{$R_{\rm p}$}}
\newcommand{\rhop}{\mbox{$\rho_{\rm p}$}}

\providecommand{\mj}{\ensuremath{\,M_{\rm J}}}
\providecommand{\rj}{\ensuremath{\,R_{\rm J}}}
\newcommand{\re}{\mbox{$R_{\ensuremath{\oplus}}$}}
\newcommand{\me}{\mbox{$M_{\ensuremath{\oplus}}$}}
\newcommand{\fe}{\mbox{$F_{\ensuremath{\oplus}}$}}

\newcommand{\kep}{\mbox{\textit{Kepler}}}
\newcommand{\target}{TOI-197}
\newcommand{\planet}{TOI-197.01}
\newcommand{\tess}{{\it TESS}}
\newcommand{\gaia}{\mbox{\textit{Gaia}}}

\newcommand{\radstar}{\mbox{$2.943\pm0.064$}}
\newcommand{\massstar}{\mbox{$1.212\pm0.074$}}
\newcommand{\agestar}{\mbox{$4.9\pm1.1$}}
\newcommand{\loggstar}{\mbox{$3.584\pm0.010$}}
\newcommand{\denstar}{\mbox{$0.06702\pm0.00067$}}

\newcommand{\teffstar}{\mbox{$5080 \pm 90$}}
\newcommand{\fehstar}{\mbox{$-0.08 \pm 0.08$}}

\newcommand{\radplanet}{\mbox{$0.836\pm0.031$}}
\newcommand{\massplanet}{\mbox{$0.190\pm0.018$}}
\newcommand{\denplanet}{\mbox{$0.431\pm0.062$}}
\newcommand{\incplanet}{\mbox{$343\pm24$}}

\newcommand{\radplanete}{\mbox{$9.17\pm0.33$}}
\newcommand{\massplanete}{\mbox{$60.5\pm5.7$}}

\shorttitle{TOI-197}  
\shortauthors{Huber et al.}

\begin{document}

\title{A HOT SATURN ORBITING AN OSCILLATING LATE SUBGIANT DISCOVERED BY TESS}

\author[0000-0001-8832-4488]{Daniel Huber}
\email{huberd@hawaii.edu}
\affiliation{Institute for Astronomy, University of Hawai`i, 2680 Woodlawn Drive, Honolulu, HI 96822, USA}

\author[0000-0002-5714-8618]{William J. Chaplin}
\affiliation{School of Physics and Astronomy, University of Birmingham, Birmingham B15 2TT, UK}
\affiliation{Stellar Astrophysics Centre (SAC), Department of Physics and Astronomy, Aarhus University, Ny Munkegade 120, DK-8000 Aarhus C, Denmark}

\author[0000-0003-1125-2564]{Ashley Chontos}
\affiliation{Institute for Astronomy, University of Hawai`i, 2680 Woodlawn Drive, Honolulu, HI 96822, USA}
\affiliation{NSF Graduate Research Fellow}

\author{Hans Kjeldsen}
\affiliation{Stellar Astrophysics Centre (SAC), Department of Physics and Astronomy, Aarhus University, Ny Munkegade 120, DK-8000 Aarhus C, Denmark}
\affiliation{Institute of Theoretical Physics and Astronomy, Vilnius University, Sauletekio av. 3, 10257 Vilnius, Lithuania}

\author[0000-0001-5137-0966]{J\o rgen Christensen-Dalsgaard}
\affiliation{Stellar Astrophysics Centre (SAC), Department of Physics and Astronomy, Aarhus University, Ny Munkegade 120, DK-8000 Aarhus C, Denmark}

\author[0000-0001-5222-4661]{Timothy R.\ Bedding}
\affiliation{Sydney Institute for Astronomy (SIfA), School of Physics, University of Sydney, NSW 2006, Australia}
\affiliation{Stellar Astrophysics Centre (SAC), Department of Physics and Astronomy, Aarhus University, Ny Munkegade 120, DK-8000 Aarhus C, Denmark}

\author[0000-0002-4773-1017]{Warrick Ball}
\affiliation{School of Physics and Astronomy, University of Birmingham, Birmingham B15 2TT, UK}
\affiliation{Stellar Astrophysics Centre (SAC), Department of Physics and Astronomy, Aarhus University, Ny Munkegade 120, DK-8000 Aarhus C, Denmark}


\author{Rafael Brahm} 
\affiliation{Center of Astro-Engineering UC, Pontificia Universidad Cat\'olica de Chile, Av. Vicu\~na Mackenna 4860, 7820436 Macul, Santiago, Chile}
\affiliation{Instituto de Astrof\'isica, Facultad de F\'isica, Pontificia Universidad Cat\'olica de Chile}
\affiliation{Millennium Institute of Astrophysics, Av. Vicu\~na Mackenna 4860, 782-0436 Macul, Santiago, Chile}

\author{Nestor Espinoza} 
\affiliation{Max-Planck-Institut fur Astronomie, K{\"o}nigstuhl 17, D-69117 Heidelberg, Germany}
\author{Thomas Henning} 
\affiliation{Max-Planck-Institut fur Astronomie, K{\"o}nigstuhl 17, D-69117 Heidelberg, Germany}
% ...

\begin{abstract}
We present the discovery of \planet, the first transiting planet identified by the \textit{Transiting Exoplanet Survey Satellite} (\textit{TESS}) for which asteroseismology of the host star is possible. \target\ (HIP\,116158) is a bright ($V=8.2$\,mag), spectroscopically classified subgiant which oscillates with an average frequency of about\,430\,\muHz\ and displays a clear signature of mixed modes. The oscillation amplitude confirms that the redder \tess\ bandpass compared to \kep\ has a small effect on the oscillations, supporting the expected yield of thousands of solar-like oscillators with \tess\ 2-minute cadence observations. Asteroseismic modeling yields a robust determination of the host star radius ($\rstar = \radstar \rsun$), mass ($\mstar = \massstar  \msun$) and age (\agestar\,Gyr), and demonstrates that it has just started ascending the red-giant branch. Combining asteroseismology with transit modeling and radial-velocity observations, we show that the planet is a ``hot Saturn'' ($\radp= \radplanete \re$) with an orbital period of $\sim$\,14.3 days, irradiance of $F=\incplanet \fe$, moderate mass ($\massp=\massplanete \me$) and density ($\rhop = \denplanet$\,g\,cm$^{-3}$). The properties of \planet\ show that the host-star metallicity -- planet mass correlation found in sub-Saturns ($4-8\re$) does not extend to larger radii, indicating that planets in the transition between sub-Saturns and Jupiters follow a relatively narrow range of densities. With a density measured to $\sim$\,15\%, \planet\ is one of the best characterized Saturn-sized planets to date, augmenting the small number of known transiting planets around evolved stars and demonstrating the power of \tess\ to characterize exoplanets and their host stars using asteroseismology.
\end{abstract}

\keywords{planets and satellites: individual (TOI-197) --- stars: fundamental parameters --- techniques: asteroseismology, photometry, spectroscopy --- TESS --- planetary systems}

\section{Introduction}
\label{sect:intro}

Asteroseismology is one of the major success stories of the space photometry revolution initiated by CoRoT \citep{baglin06b} and \kep\ \citep{borucki10}. The detection of oscillations in thousands of stars has led to breakthroughs such as the discovery of rapidly rotating cores in subgiants and red giants, as well as the systematic measurement of stellar masses, radii and ages \citep[see][for a review]{chaplin13a}. Asteroseismology has also become the ``gold standard'' for calibrating more indirect methods to determine stellar parameters such as surface gravity (\logg) from spectroscopy \citep{petigura17b} and stellar granulation \citep{mathur11b,bastien13,kallinger16,corsaro17,bugnet2018,pande2018}, and age from rotation periods \citep[gyrochronology, e.g.][]{garcia14,vansaders16}. 

A remarkable synergy that emerged from space-based photometry is the systematic characterization of exoplanet host stars using asteroseismology.  Following first asteroseismic studies of exoplanet host stars using radial velocities \citep{bouchy05,bazot05}, the Hubble Space Telescope \citep{gilliland11} and CoRoT \citep{ballot11b,lebreton14}, \kep\ enabled the systematic characterization of exoplanets with over 100 detections of oscillations in host stars to date \citep{huber13,lundkvist16}. In addition to the more precise characterization of exoplanet radii and masses \citep{ballard14}, the synergy also enabled systematic constraints on stellar spin-orbit alignments \citep{Chaplin13,benomar14,lund14,campante16b} and statistical inferences on orbital eccentricities through constraints on the mean stellar density \citep{sliski14,vaneylen15,vaneylen18}.

The recently launched NASA \tess\ Mission \citep{ricker14} is poised to continue the synergy between asteroseismology and exoplanet science. Using dedicated 2-minute cadence observations, \tess\ is expected to detect oscillations in thousands of main-sequence, subgiant and early red-giant stars \citep{schofield18}, and simulations predict that at least 100 of these will host transiting or non-transiting exoplanets \citep{campante16}. \tess\ host stars are on average significantly brighter than typical \kep\ hosts, facilitating ground-based measurements of planet masses with precisely characterized exoplanet hosts from asteroseismology. 
While some of the first exoplanets discovered with \tess\ orbit stars that have evolved off the main sequence \citep{wang:2018, brahm:2018, nielsen:2018}, none of them were amenable to asteroseismology using \tess\ photometry. 
Here, we present the characterization of TESS Object of Interest 197 (TOI-197, HIP\,116158) system, the first discovery by \tess\ of a transiting exoplanet around a host star in which oscillations can be measured.

\section{Observations}

\subsection{TESS Photometry}

\begin{figure*}[ht!]
\begin{center}
\resizebox{16cm}{!}{\includegraphics{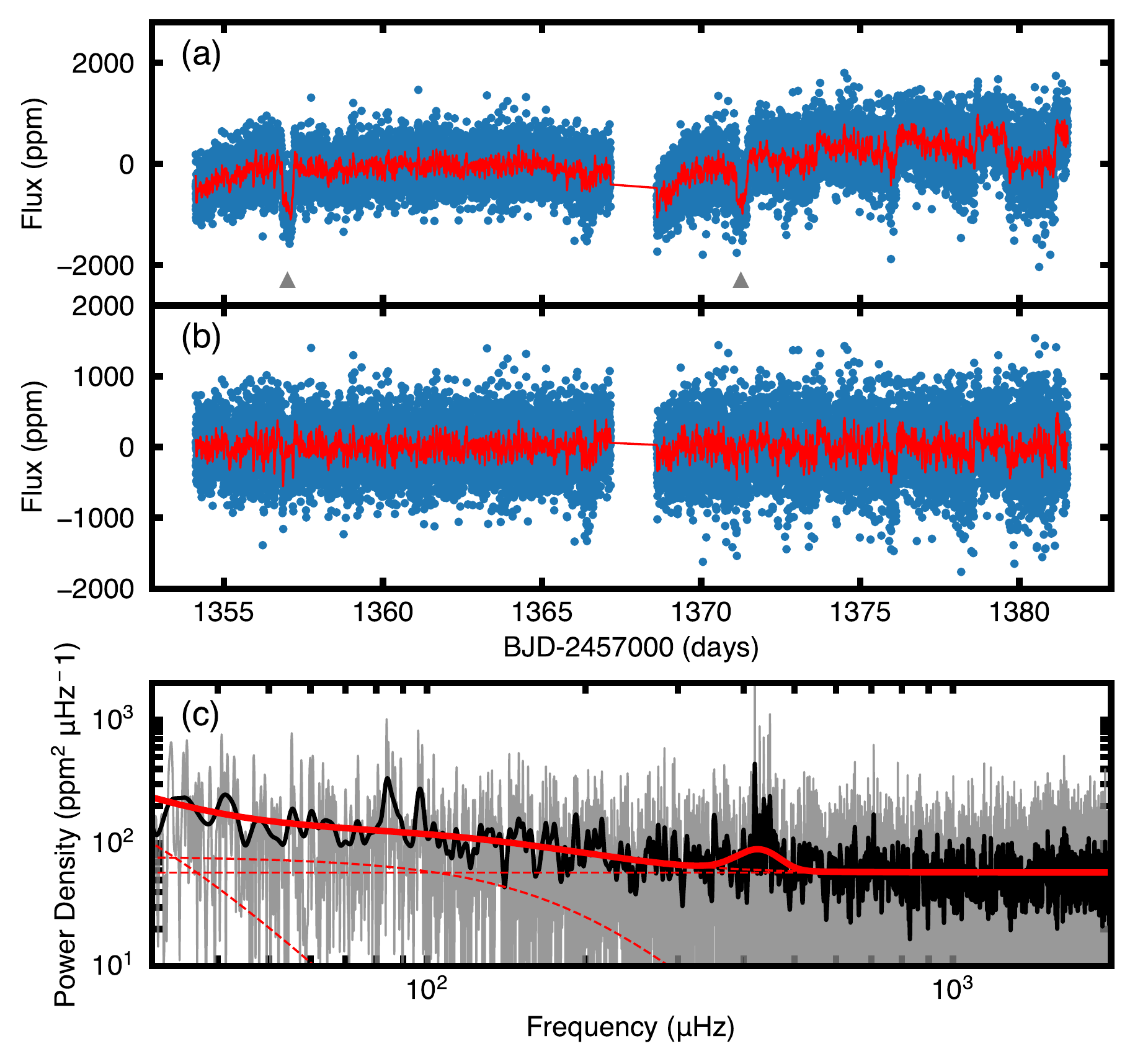}}
\caption{Panel (a): Raw \tess\ 2-minute cadence light curve of \target\ produced by the TESS Asteroseismic Science Operations Center (TASOC). The red line is the light curve smoothed with a 10-minute boxcar filter (shown for illustration purposes only). Upward triangles mark the two transit events. Panel (b): Light curve after applying corrections by the TASOC pipeline. Panel (c): Power spectrum of panel (b), showing a granulation background and power excess due to oscillations near $\sim$\,430\,\muHz. The solid red line is a global fit, consisting of granulation plus white noise and a Gaussian describing the power excess due to oscillations. Dashed red lines show the two granulation components and the white noise level, respectively.}
\label{fig:ts}
\end{center}
\end{figure*}

\tess\ observed \target\ in 2-minute cadence during Sector 2 of Cycle 1 for 27 days. We used the target pixel files produced by the TESS Science Processing Operations Center \citep{jenkins16} as part of the TESS alerts on November 11 2018\footnote{\dataset[https://doi.org/10.17909/t9-wx1n-aw08]{https://doi.org/10.17909/t9-wx1n-aw08}}. We produced a light curve using the photometry pipeline\footnote{\url{https://tasoc.dk/code/}} (Handberg et al., in prep.) maintained by the TESS Asteroseismic Science Operations Center 
\citep[TASOC,][]{lund17}, which is based on software originally developed to generate light curves for data collected by the K2 Mission \citep{lund15}.

Figure \ref{fig:ts}a shows the raw light curve obtained from the TASOC pipeline. The coverage is nearly continuous (duty cycle $\sim$\,93\%), with a $\sim$\,2\,day gap separating the two spacecraft orbits in the observing sector. Two $\sim$\,0.1\,\% brightness dips, which triggered the identification of \planet\ as a planet candidate, are evident near the beginning of each \tess\ orbit (see upward triangles in Figure \ref{fig:ts}a). The structure with a period of $\sim$\,2.5\,d corresponds to instrumental variations due to the angular momentum dumping cycle of the spacecraft.

To prepare the raw light curve for an asteroseismic analysis, the current TASOC pipeline implements a series of corrections as described by \citet{handberg14}, which includes removal of instrumental artefacts and of the transit events using a combination of filters utilizing the estimated planetary period. Future TASOC-prepared light curves from full \tess\ data releases will use information from the ensemble of stars to remove common instrumental systematics (Lund et al, in prep.). 
Alternative light curve corrections using transit removal and gap interpolation \citep{garcia11,pires15} yielded consistent results.
The corrected TASOC light curve is shown in Figure \ref{fig:ts}b. Figure \ref{fig:ts}c shows a power spectrum of this light curve, revealing the clear presence of a granulation background and a power excess from solar-like oscillations near $\sim$\,430\,\muHz, both characteristic of an evolved star near the base of the red-giant branch.

\subsection{High-Resolution Spectroscopy}
\label{sec:spectroscopy}

We obtained high-resolution spectra of \target\ using several facilities within the TESS Follow-up Observation Program (TFOP), including 
HIRES \citep{vogt94} on the 10-m telescope at Keck Observatory (Maunakea, Hawai`i), 
the Hertzsprung SONG Telescope at Teide Observatory (Tenerife) \citep{grundahl17}, 
HARPS \citep{mayor03}, FEROS \citep{kaufer99}, Coralie \citep{queloz01} and FIDEOS \citep{vanzi:2018} on the MPG/ESO 3.6-m, 2.2-m, 1.2-m, and 1-m telescopes at La Silla Observatory (Chile), 
Veloce \citep{veloce} on the 3.9-m Anglo-Australian Telescope at Siding Spring Observatory (Australia),
TRES \citep{furesz08} on the 1.5-m Tillinghast reflector at the F. L. Whipple Observatory (Mt. Hopkins, Arizona), and iSHELL \citep{rayner12} on the NASA IRTF Telescope (Maunakea, Hawaii). 
All spectra used in this paper were obtained between Nov 11 and Dec 30 2018 and have a minimum spectral resolution of $R\approx 44000$. FEROS, Coralie, and HARPS data were processed and analyzed with the CERES package \citep{brahm:2017}, which performs the optimal extraction and wavelength calibration of each spectrum, along with the measurement of precision radial velocities and bisector spans via the cross-correlation technique. Most instruments have been previously used to obtain precise radial velocities to confirm exoplanets, and we refer to the publications listed above for details on the reduction methods.

To obtain stellar parameters, we analyzed a HIRES spectrum  using Specmatch \citep{petigura15b}, which has been extensively applied for the classification of \kep\ exoplanet host stars \citep{johnson17,petigura17b}. The resulting parameters were $\teff=5080\pm70$\,K, $\logg=3.60\pm0.08$\,dex, $\feh=-0.08\pm0.05$\,dex and $v\sin{i}=2.8 \pm 1.6$\,km\,s$^{-1}$, consistent with an evolved star as identified from the power spectrum in Figure \ref{fig:ts}c. To account for systematic differences between spectroscopic methods \citep{torres12} we added 59\,K in \teff\ and 0.062\,dex in \feh\ in quadrature to the formal uncertainties, yielding final values of $\teff=5080\pm90$\,K and $\feh=-0.08\pm0.08$\,dex. Independent spectroscopic analyses yielded consistent results, including an analysis of a HIRES spectrum using ARES+MOOG \citep{sousa14,sousa18}, FEROS spectra using ZASPE \citep{brahm:2017b}, TRES spectra using SPC \citep{buchhave12} and iSHELL spectra using BT-Settl models \citep{allard12}.


\subsection{Broadband Photometry \& Gaia Parallax}
\label{sec:sed}

We fitted the spectral energy distribution (SED) of \target\ using broadband photometry following the method described by \citet{stassun16a}. We used NUV photometry from {\it GALEX}, $B_T V_T$ from {\it Tycho-2} \citep{hog00}, $BVgri$ from APASS, $JHK_S$ from {\it 2MASS} \citep{skrutskie06}, W1--W4 from {\it WISE} \citep{wright10}, and the $G$ magnitude from {\it Gaia} \citep{evans18}. The data were fit using Kurucz atmosphere models, with \teff, \feh\ and extinction ($A_V$) as free parameters. We restricted $A_V$ to the maximum line-of-sight value from the dust maps of \citet{schlegel98}. The resulting fit yielded $T_{\rm eff} = 5090 \pm 85$~K, ${\rm [Fe/H]} = -0.3 \pm 0.3$\,dex, and $A_V = 0.09 \pm 0.02$\,mag with reduced $\chi^2$ of 1.9, in good agreement with spectroscopy. Integrating the (de-reddened) model SED gives the bolometric flux at Earth of $F_{\rm bol} = 1.88 \pm 0.04 \times 10^{-8}$ erg~s~cm$^{-2}$. An independent SED fit using 2MASS, APASS9, USNO-B1 and WISE photometry and Kurucz models yielded excellent agreement, with $F_{\rm bol} = 1.83 \pm 0.09 \times 10^{-8}$ erg~s~cm$^{-2}$ and $\teff=5150 \pm 130$\,K. Additional independent analyses using the method by \citet{mann16} and PARAM \citep{rodrigues14,rodrigues17} yielded bolometric fluxes and extinction values that are consistent within 1\,$\sigma$ with the values quoted above.

Combining the bolometric flux with the \gaia\ DR2 distance allows us to derive a nearly model-independent luminosity, which is a valuable constraint for asteroseismic modeling (see Section \ref{sec:modeling}). Using a \gaia\ parallax of $10.518 \pm 0.080$\,mas (adjusted for the $0.082 \pm 0.033$~mas zero-point offset for nearby stars reported by \citealt{stassun18b}) with the two methods described above yielded $\lstar=5.30 \pm 0.14\lsun$ (using $F_{\rm bol} = 1.88 \pm 0.04 \times 10^{-8}$ erg~s~cm$^{-2}$) and $\lstar=5.13 \pm 0.13\lsun$ (using $F_{\rm bol} = 1.83 \pm 0.09 \times 10^{-8}$ erg~s~cm$^{-2}$). We also derived a luminosity using \texttt{isoclassify} \citep{huber17}\footnote{\url{https://github.com/danxhuber/isoclassify}}, adopting 2MASS $K$-band photometry, bolometric corrections from MIST isochrones \citep{choi16} and the composite reddening map \texttt{mwdust} \citep{bovy16}, yielding $\lstar=5.03 \pm 0.13\lsun$. Our adopted luminosity was the mean of these methods with an uncertainty calculated by adding the mean uncertainty and scatter over all methods in quadrature, yielding  $\lstar=5.15\pm0.17\lsun$.

\subsection{High-Resolution Imaging}
\label{sec:ao}

\target\ was observed with the NIRC2 camera and Altair adaptive optics system on Keck-2 \citep{wizinowich00} on UT 25 November 2018.  Conditions were clear but seeing was poor (0.8--2'').  We used the science target as the natural guide star and images were obtained through a $K$-continuum plus KP50$_{1.5}$ filter using the narrow camera (10 mas pixel scale).  We obtained eight images (four each at two dither positions), each consisting of 50 co-adds of 0.2~sec each, with correlated double-sampling mode and four reads.  Frames were co-added and we subtracted an average dark image, constructed from a set of darks with the same integration time and sampling mode.  Flat-fielding was performed using a dome flat obtained in the $K'$ filter.  ``Hot'' pixels were identified in the dark image and corrected by median filtering with a $5 \times 5$ box centered on each affected pixel in the science image. Only a single star appears in the images.  We performed tests in which ``clones'' of the stellar image reduced by a specified contrast ratio were added to the original image.  These show that we would have been able to detect companions as faint as $\Delta K = 5.8$\,mag within 0.4'' of \target, 3.8\,mag within 0.2'', and 1.8\,mag within 0.1''. 

Additional NIRC2 observations were obtained in the narrow-band $Br-\gamma$ filter $(\lambda_o = 2.1686; \Delta\lambda = 0.0326\mu$m) on UT 22 November 2018. A standard 3-point dither pattern with a step size of $3\arcsec$ was repeated twice with each dither offset from the previous dither by $0.5\arcsec$. An integration time of 0.25 seconds was used with one coadd per frame for a total of 2.25 seconds on target, and the camera was used in the narrow-angle mode. No additional stellar companions were detected to within a resolution of $\sim 0.05\arcsec$ FWHM. The sensitivities of the final combined AO image were determined following \citet{ciardi15} and \citet{furlan17}, with detection limits as faint as $\Delta Br-\gamma = 7.4$\,mag within 0.4'', 6.1\,mag within 0.2'', and 3.2\,mag within 0.1''.

The results from NIRC2 are consistent with Speckle observations using HRCam \citep{tokovinin10} on the 4.1\,m SOAR telescope\footnote{\url{https://exofop.ipac.caltech.edu/tess/target.php?id=441462736}}. Since the companion is unlikely to be bluer than TOI-197, these constraints exclude any significant dilution (both for oscillation amplitudes and the depth of transit events).

\section{Asteroseismology}

\subsection{Global Oscillation Parameters}
\label{sec:global}

\begin{figure}
\begin{center}
\resizebox{\hsize}{!}{\includegraphics{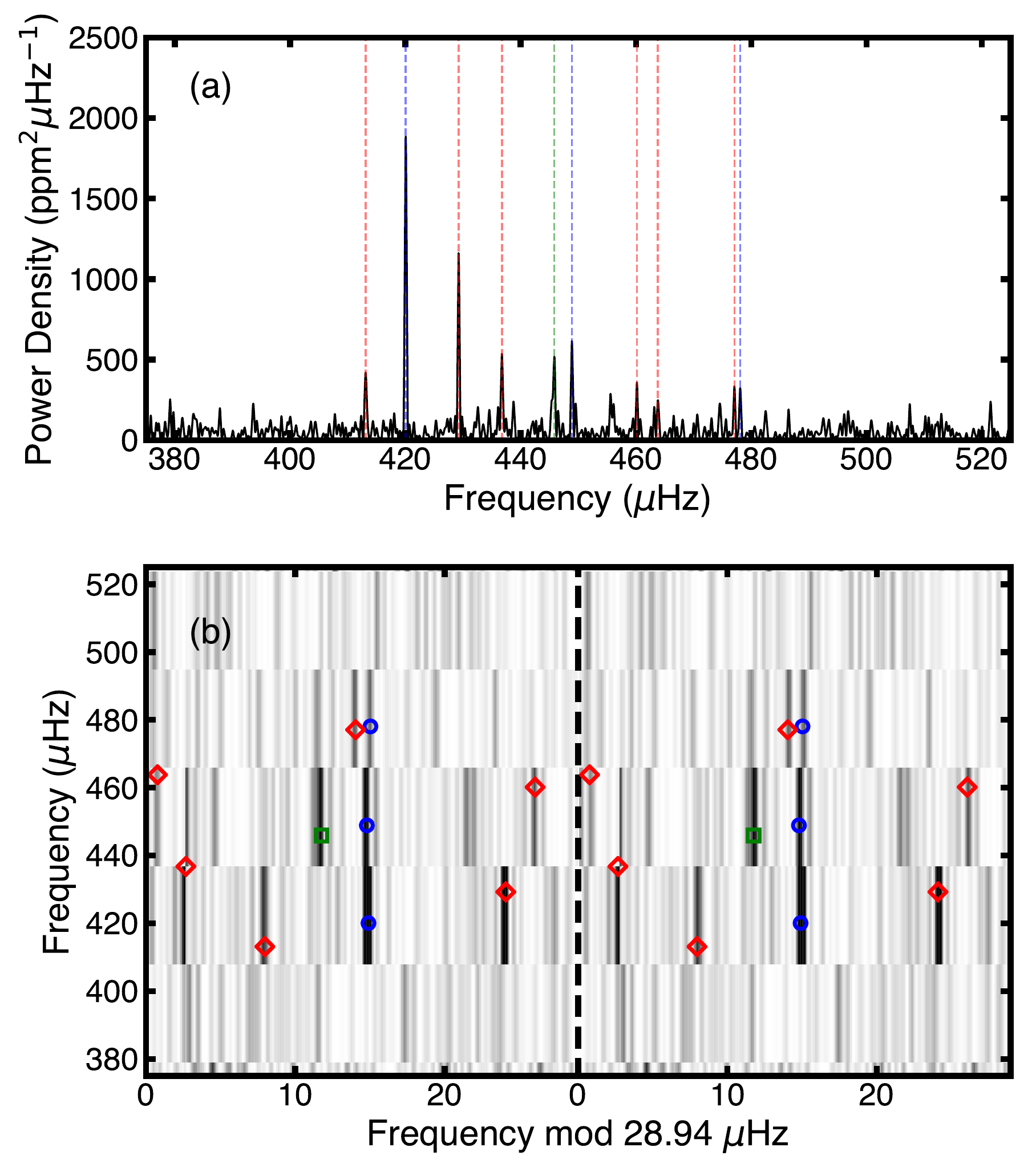}}
\caption{Panel (a): Power spectrum of \target\, 
centered on the frequency region showing oscillations. Vertical dashed lines mark identified individual frequencies.
Panel (b): Greyscale \'{e}chelle diagram$^{5}$ of the background-corrected and smoothed power spectrum in panel (a). Identified individual mode frequencies are marked with blue circles ($l=0$, radial modes), green squares ($l=2$, quadrupole modes) and red diamonds ($l=1$, dipole modes). Note that the diagram is replicated for clarity \citep{bedding12}.}
\label{fig:echelle}
\end{center}
\end{figure}

To extract oscillation parameters characterizing the average properties of the power spectrum, we used several automated analysis methods \citep[e.g.][]{huber09,mathur10b,mosser11c,benomar12,kallinger12,corsaro14,ref:lundkvist2015,stello17,campante18,bell19}, many of which have been extensively tested on \kep\ data \citep[e.g.][]{hekker11,verner11}.  In most of these analyses, the power contributions due to granulation noise and stellar activity were modeled by a combination of power laws and a flat contribution due to shot noise, and then corrected by dividing the power spectrum by the background model. The individual contributions and background model using the method by \citet{huber09} are shown as dashed and solid red lines in Fig.~\ref{fig:ts}c, and a close-up of the power excess is shown in Fig.~\ref{fig:echelle}a.

Next, the frequency of maximum power (\numax) was measured either by heavily smoothing the power spectrum or by fitting a Gaussian function to the power excess. Our analysis yielded $\numax=430\pm18\,\muHz$, with uncertainties calculated from the scatter between all fitting techniques. Finally, the mean oscillation amplitude per radial mode was determined by taking the peak of the smoothed, background-corrected oscillation envelope and correcting for the contribution of non-radial modes \citep{kjeldsen08}, yielding $A=18.7\pm3.5$\,ppm. We caution that the \numax\ and amplitude estimates could be significantly biased by the stochastic nature of the oscillations. The modes are not well resolved, as demonstrated by the non-Gaussian appearance of the power spectrum and the particularly strong peak at 420\,\muHz. 

\begin{figure}
\begin{center}
\resizebox{\hsize}{!}{\includegraphics{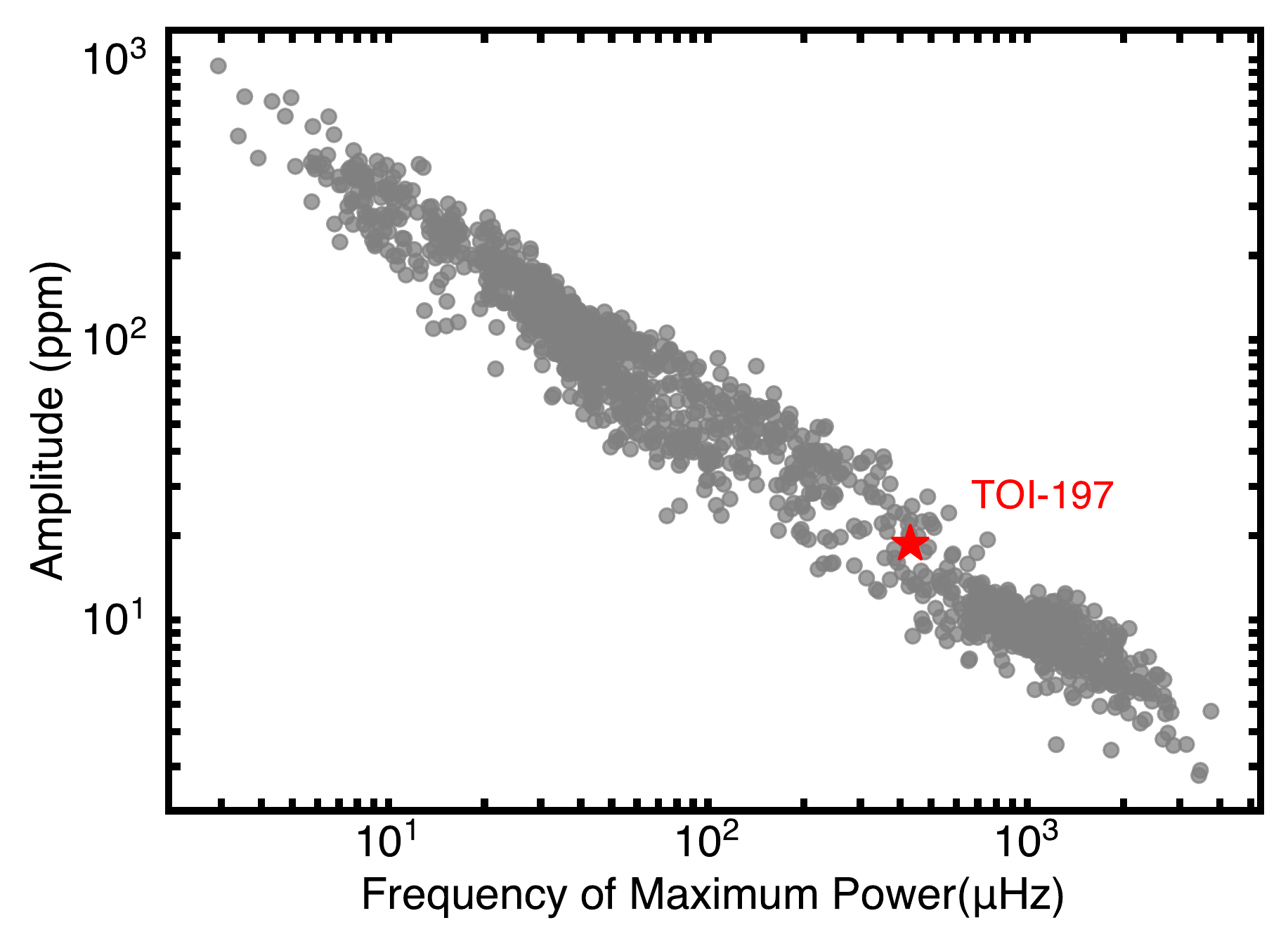}}
\caption{Amplitude per radial mode versus frequency of maximum power for a sample of $\sim$\,1500 stars spanning from the main-sequence to the red giant branch observed by \kep\ \citep{huber11}. The red star shows the measured position of TOI-197. The uncertainties are approximately equal to the symbol size.}
\label{fig:amplitude}
\end{center}
\end{figure}

\let\thefootnote\relax\footnote[5]{\noindent $^{5}$\'{E}chelle diagrams are constructed by dividing a power spectrum into equal segments with length~\dnu\ and stacking one above the other, so that modes with a given spherical degree align vertically in ridges \citep{grec83}. Departures from regularity arise from sound speed discontinuities and from mixed modes, and thus probe the interior structure of a star.}
Global seismic parameters such as \numax\ and amplitude follow well-known scaling relations \citep{huber11,mosser12,corsaro13}, which allow us to test whether the detected oscillations are consistent with expectations. Figure~\ref{fig:amplitude} compares our measured \numax\ and amplitude with results for $\sim$1500 stars observed by \kep\ \citep{huber11}. We observe excellent agreement, confirming that the detected signal is consistent with solar-like oscillations. We note that the oscillations in the \tess\ bandpass are expected to be $\sim$\,15\,\% smaller than in the bluer \kep\ bandpass, which is well within the spread of amplitudes at a given \numax\ observed in the \kep\ sample. The result confirms that the redder bandpass of \tess\ only has a small effect on the oscillation amplitude, supporting the expected rich yield of solar-like oscillators with \tess\ 2-minute cadence observations \citep{schofield18}.

\subsection{Individual Mode Frequencies}

The power spectrum in Fig.~\ref{fig:echelle}a shows several clear peaks corresponding to individual oscillation modes. 
Given that \tess\ instrument artifacts are not yet well understood, we restricted our analysis to the frequency range 400--500\,\muHz\ where we observe peaks well above the background level.

To extract these individual mode frequencies, we used several independent methods ranging from traditional iterative sine-wave fitting, i.e., pre-whitening \citep[e.g.][]{lenz05,kjeldsen05,bedding07}, to fitting of Lorentzian mode profiles \citep[e.g.][]{handberg11,appourchaux12,mosser12,corsaro14, corsaro15,vrard15,davies16,roxburgh17,handberg17,kallinger18}, including publicly available code such as \texttt{DIAMONDS}$^{6}$\footnote{$^{6}$\url{https://github.com/EnricoCorsaro/DIAMONDS}}. We required at least two independent methods to return the same frequency within uncertainties and that the posterior probability of each peak being a mode was $\ge 90\,\%$ \citep{basuchaplin17}. A comparison of the frequencies returned by different fitters showed very good agreement, at a level smaller than the uncertainties for all the reported modes. For the final list of frequencies we adopted values from one fitter who applied pre-whitening (HK), with uncertainties derived from Monte Carlo simulations of the data, as listed in Table~\ref{tab:freqs}. 

To measure the large frequency separation \dnu, we performed a linear fit to all identified radial modes, yielding $\dnu=28.94 \pm 0.15 \muHz$. Figure~\ref{fig:echelle}b shows a greyscale \'{e}chelle diagram$^{5}$ using this \dnu\ measurement, including the extracted mode frequencies. The $l=1$ modes are strongly affected by mode bumping, as expected for the mixed mode coupling factors for evolved stars in this evolutionary stage. The offset $\epsilon$ of the $l=0$ ridge is $\sim\,1.5$, consistent with the expected value from \kep\ measurements for stars with similar $\dnu$ and $\teff$ \citep{white11}.

\begin{table}
\begin{center}
\caption{Extracted oscillation frequencies and mode identifications for \target.}
\vspace{0.1cm}
\begin{tabular}{c c c}        
$f(\muHz)$  & $\sigma_{f} (\muHz)$ & $l$ \\
\hline         
   413.12  &	  0.29  &	1  \\
   420.06  &	  0.11  &	0  \\
   429.26  &	  0.14  &	1  \\
   436.77  &	  0.24  &	1  \\
   445.85  &	  0.21  &	2  \\
   448.89  &	  0.21  &	0  \\
   460.16  &	  0.33  &	1  \\
   463.81  &	  0.43  &	1  \\
   477.08  &	  0.31	&	1  \\
   478.07  &	  0.35	&	0  \\
\hline         
\end{tabular}
\label{tab:freqs}
\end{center}
\flushleft Note: The large frequency separation derived from radial modes is $\dnu=28.94 \pm 0.15 \muHz$. Note that the $l=1$ modes at $\sim$\,460 and $\sim$\,463\,\muHz\ are listed for completeness, but it is unlikely that both of them are genuine (see text).
\end{table}

\subsection{Frequency Modeling}
\label{sec:modeling}

\begin{figure}
\begin{center}
\resizebox{\hsize}{!}{\includegraphics{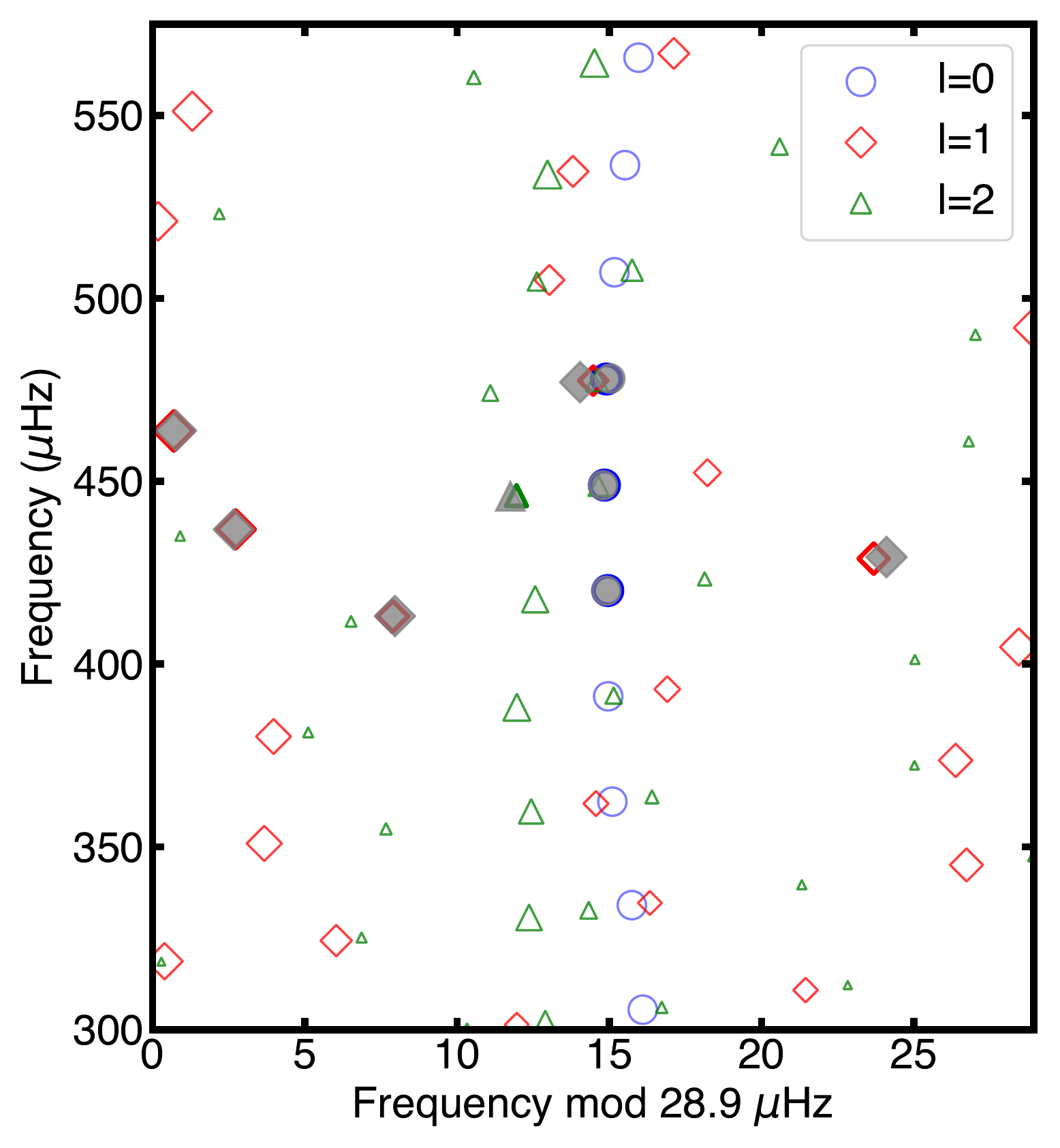}}
\caption{\'{E}chelle diagram showing observed oscillation frequencies (filled grey symbols) and a representative best-fitting model (open colored symbols) using GARSTEC, ADIPLS and BeSSP \citep{serenelli17}. Model symbol sizes for non-radial modes are scaled using the mode inertia (a proxy for mode amplitude) as described in \citet{cunha15}. Thick model symbols correspond to modes that were matched to observations. Uncertainties on the observed frequencies are than smaller or comparable to the symbol sizes. Note that the $l=1$ mode at 460\,\muHz\ has been omitted from this plot (see text).}
\label{fig:modelechelle}
\end{center}
\end{figure} 

We used a number of independent approaches to model the observed oscillation frequencies, including different stellar evolution codes \citep[ASTEC, Cesam2K, GARSTEC, Iben, MESA, and YREC,][]{jcd08,morel08,scuflaire08,weiss08,iben65,paxton11,paxton13,paxton15,choi16,demarque08}, oscillation codes \citep[ADIPLS, GYRE and Pesnell,][]{jcd08,gyre,pesnell90} and modeling methods \citep[including AMP, ASTFIT, BeSSP, BASTA, PARAM,][]{creevey17,silva15,serenelli17,rodrigues14,rodrigues17,deheuvels11,yildiz16,ong18,tayar18,lebreton14,ball17,mosumgaard18}.
Most of the adopted methods applied corrections for the surface effect \citep{kjeldsen08b,ball17}.
Model inputs included the spectroscopic temperature and metallicity, individual frequencies, \dnu, and the luminosity (Section \ref{sec:sed}). To investigate the effects of different input parameters, modelers were asked to provide solutions using both individual frequencies and only using \dnu, with and without taking into account the luminosity constraint. The constraint on \numax\ was not used in the modeling since it may be affected by finite mode lifetimes (see Section \ref{sec:global}).

Overall, the modeling efforts yielded consistent results, and most modeling codes were able to provide adequate fits to the observed oscillation frequencies. The modeling confirmed that only one of the two closely-spaced mixed modes near $\sim$\,460\,\muHz\ is likely real, but we have retained both frequencies in Table \ref{tab:freqs} for consistency. An \'{e}chelle diagram with observed frequencies and a representative best-fitting model is shown in Figure \ref{fig:modelechelle}.

\begin{table}
\begin{center}
\caption{Host Star Parameters}
\renewcommand{\tabcolsep}{0mm}
\begin{tabular}{l c}
\tableline\tableline
\noalign{\smallskip}
\multicolumn{2}{c}{Basic Properties} \\
\noalign{\smallskip}
\hline
\noalign{\smallskip}
Hipparcos ID & 116158 \\
TIC ID & 441462736 \\
$V$ Magnitude & 8.15 \\
\tess\ Magnitude & 7.30  \\
$K$ Magnitude & 6.04 \\
\noalign{\smallskip}
\hline
\noalign{\smallskip}
\multicolumn{2}{c}{SED \& \gaia\ Parallax} \\
\noalign{\smallskip}
\hline
\noalign{\smallskip}
Parallax, $\pi$ (mas) & $10.518 \pm 0.080$  \\
Luminosity, $L$ ($\lsun$) & $5.15 \pm 0.17$ \\
\noalign{\smallskip}
\hline
\noalign{\smallskip}
\multicolumn{2}{c}{Spectroscopy} \\
\noalign{\smallskip}
\hline
\noalign{\smallskip}
Effective Temperature, \teff\, (K) & \teffstar \\
Metallicity, [Fe/H] (dex) & \fehstar \\
Projected rotation speed, \vsini\ (km\,s$^{-1}$) & $2.8 \pm 1.6$ \\
\noalign{\smallskip}
\hline
\noalign{\smallskip}
\multicolumn{2}{c}{Asteroseismology} \\
\noalign{\smallskip}
\hline
\noalign{\smallskip}
Stellar Mass, \mstar\ (\msun)& \massstar \\
Stellar Radius, \rstar\ (\rsun)& \radstar \\
Stellar Density, \rhostar\ (gcc)& \denstar \\
Surface gravity, \logg\ (cgs) & \loggstar \\
Age, $t$ (Gyr) & \agestar \\
\noalign{\smallskip}
\hline
\end{tabular}
\end{center}
\flushleft Notes: The \tess\ magnitude is adopted from the TESS Input Catalog \citep{tic}.
\label{tab:stellar}
\end{table}

Independent analyses confirmed a bimodality splitting into lower-mass, older models ($\sim\,1.15\msun$, $\sim$\,6\,Gyr) and higher-mass, younger models ($\sim\,1.3\msun$, $\sim$\,4\,Gyr). Surface rotation would provide an independent mass diagnostic \citep[e.g.][]{vansaders13}, but the insufficiently constrained \vsini\ and the unknown stellar inclination mean that we cannot decisively break this degeneracy. 
Combining an independent constraint of $\logg = 3.603\pm0.026$\,dex from an autocorrelation analysis of the light curve \citep{kallinger16} with a radius from $L$ and $\teff$ favors a higher-mass solution ($\mstar = 1.27 \pm 0.13 \msun$), but may be prone to small systematics in the $\numax$ scaling relation (which was used for the calibration).
To make use of the most observational constraints available, we used the set of nine modeling solutions which used \teff, \feh, frequencies and the luminosity as input parameters. From this set of solutions, we adopted the self-consistent set of stellar parameters with the mass closest to the median mass over all results. A more detailed study of the individual modeling results will be presented in a follow-up paper (Li et al., in prep).

For ease of propagating stellar parameters to exoplanet modeling (see next section), uncertainties were calculated by adding the median uncertainty for a given stellar parameter in quadrature to the standard deviation of the parameter for all methods. This method has been commonly adopted for \kep\ \citep[e.g.][]{chaplin14} and captures both random and systematic errors estimated from the spread among different methods. For completeness, the individual random and systematic error estimates are $\rstar = 2.943 \pm 0.041 \rm{(ran)} \pm 0.049 \rm{(sys)}\,\rsun$, $\mstar = 1.212 \pm 0.052 \rm{(ran)} \pm 0.055 \rm{(sys)}\, \msun$, $\rhostar = 0.06702 \pm 0.00019 \rm{(ran)} \pm 0.00047 \rm{(sys)\,gcc}$, and $t = 4.9 \pm 0.6 \rm{(ran)} \pm 0.9 \rm{(sys)}$\,Gyr. This demonstrates that systematic errors constitute a significant fraction of the error budget for all stellar properties (in particular stellar age), and emphasize the need for using multiple model grids to derive realistic uncertainties for stars and exoplanets. The final estimates of stellar parameters are summarized in Table \ref{tab:stellar}, constraining the radius, mass, density and age of \target\ to $\sim$\,2\,\%, $\sim$\,6\,\%, $\sim$\,1\,\% and $\sim$\,22\,\%.

\section{Planet Characterization}

\begin{figure*}
\begin{center}
\resizebox{\hsize}{!}{\includegraphics{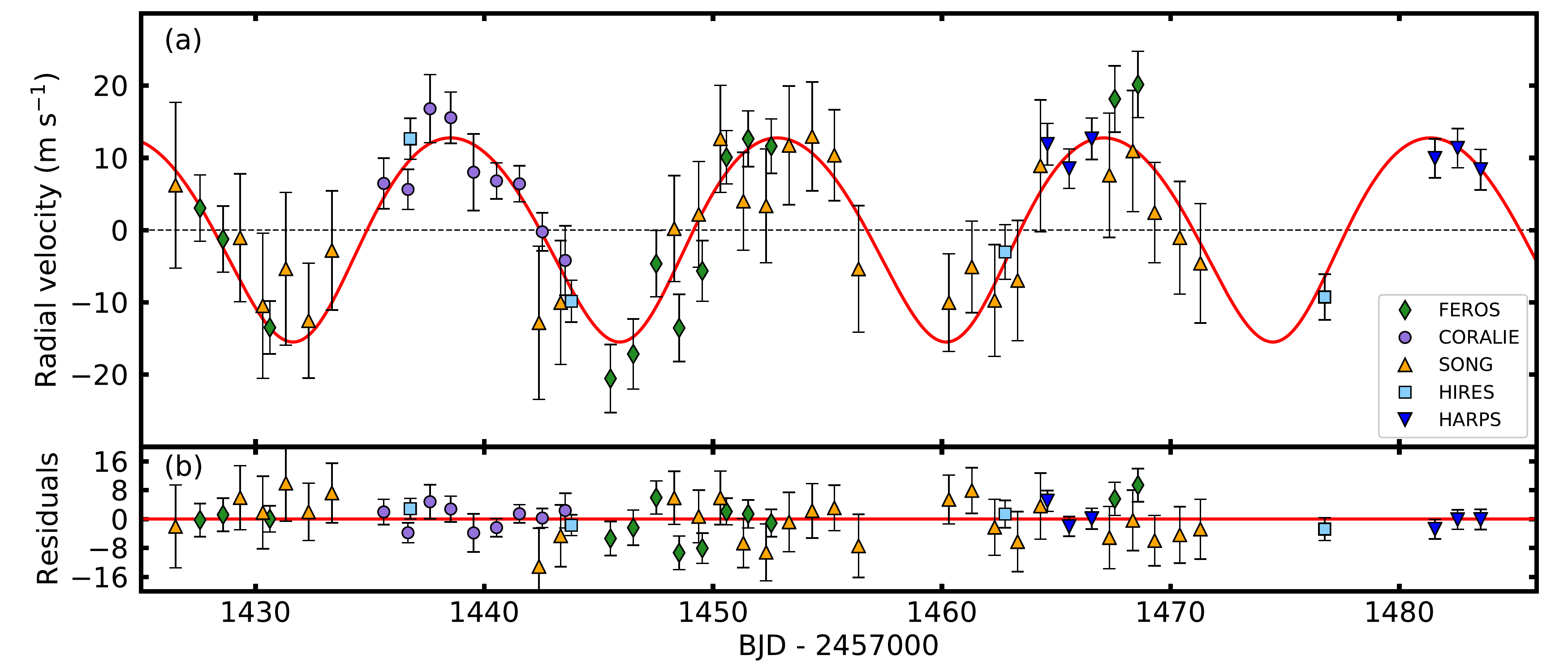}}
\caption{Radial velocity timeseries (panel a) and residuals after subtracting the best-fitting model (panel b) for \target. Datapoints are corrected for zeropoint offsets of individual instruments, and error bars include contributions from stellar jitter.}
\label{fig:rvs}
\end{center}
\end{figure*}

\begin{table}
\begin{center}
\caption{High-precision Radial Velocities for \target}
\begin{tabular}{c c c c}        
\hline         
Time (BJD) & RV (m/s) &  $\sigma_{\text{RV}}$ (m/s) & Instrument \\
\hline  
2458426.334584 & 4.258 & 11.260 & SONG \\
2458426.503655 & 6.328 & 11.270 & SONG \\
2458427.575230 & -12.667 & 3.000 & FEROS \\
2458428.547576 & 17.328 & 18.540 & SONG \\
\ldots & \ldots & \ldots & \ldots \\
2458443.535340 & -14.667 & 3.600 & CORALIE \\
2458443.541210 & -3.067 & 3.800 & CORALIE \\
2458443.714865 & -6.815 & 0.780 & HIRES \\
2458443.825283 & -4.375 & 0.720 & HIRES \\
\ldots & \ldots & \ldots & \ldots \\
2458482.562290 & 19.433 & 2.000 & HARPS \\
2458483.541710 & 16.133 & 2.000 & HARPS \\
2458483.553240 & 19.233 & 2.000 & HARPS \\
2458483.564690 & 16.233 & 2.000 & HARPS \\ 
\hline
\end{tabular} 
\label{tab:rvs} 
\end{center}
\flushleft Notes: Error bars do not include contributions from stellar jitter and measurements have not been corrected for zeropoint offsets. This table is available in its entirety in a machine-readable form in the online journal. 
\end{table}

To fit the transits observed in the \tess\ data we used the PDC-MAP light curve provided by the TESS Science Processing and Operations Center (SPOC), which has been optimized to remove instrumental variability and preserve transits \citep{smith12,stumpe14}. To optimize computation time we discarded all data more than 2.5 days before and after each of the two observed transits. We have repeated the fit and data preparation procedure using the TASOC light curve and found consistent results.

\begin{figure}[h!]
\begin{center}
\resizebox{\hsize}{!}{\includegraphics{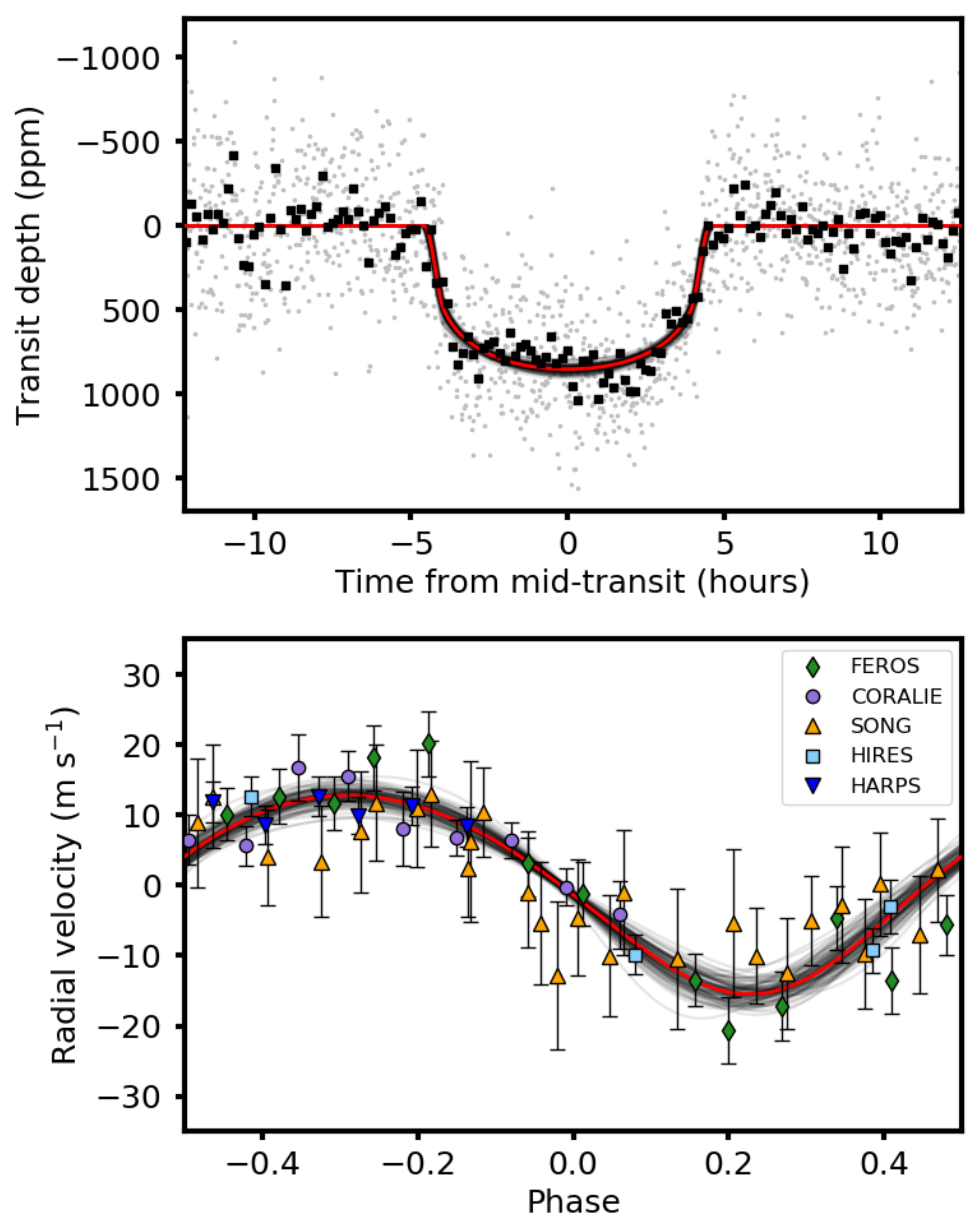}}
\caption{\tess\ light curve (panel a) and radial-velocity measurements (panel b) folded with the best fitting orbital period. Grey points in panel a show the original sampling, and black points are binned means over 10 minutes. Red lines in both pabels show the best-fitting model from the joint fit using stellar parameters, transit and radial velocities. Grey lines show random draws from the joint MCMC model. Error bars in panel b include contributions from stellar jitter.}
\label{fig:planet}
\end{center}
\end{figure}

A total of 107 radial velocity measurements from five different instruments (see Section \ref{sec:spectroscopy} and Table \ref{tab:rvs}) were used to constrain the mass of the planet. 
No spectroscopic observations were taken during transits, and hence the measurements are unaffected by the Rossiter-McLaughlin effect ($\sim$\,2.3\,m\,s$^{-1}$ based on the measured \vsini\ and $R_{p}/R_\star$).
To remove variations from stellar oscillations, we calculated weighted nightly means for all instruments which obtained multiple observations per night. We performed a joint transit and radial-velocity fit using a Markov Chain Monte Carlo algorithm based on the exoplanet modeling code \texttt{ktransit} \citep{ktransit}, as described in \citet{chontos19}. We placed a strong Gaussian prior on the mean stellar density using the value derived from asteroseismology (Table \ref{tab:stellar}) and weak priors on the linear and quadratic limb darkening coefficients, derived from the closest $I$-band grid points in \citet{claret11}, with a width of 0.6 for both coefficients. We also adopted a prior for the radial-velocity jitter from granulation and oscillations of $2.5 \pm 1.5$\,m\,s$^{-1}$, following \citet{yu18} \citep[see also][]{tayar18b}, and a $1/e$ prior on the eccentricity to account for the linear bias introduced by sampling in $e\cos{\omega}$ and $e\sin{\omega}$ \citep{eastman13}. Independent zeropoint offsets and stellar jitter values for each of the five instruments that provided radial velocities. Independent joint fits using EXOFASTv2 \citep{eastman13} yielded consistent results.

\begin{table}
\caption{Planet Parameters}
\centering
\begin{tabular}{L{1.7cm} C{1.25cm} C{1.25cm} C{1.25cm} C{1.25cm}}
\noalign{\smallskip}
\tableline\tableline
\noalign{\smallskip}
\textbf{Parameter} &  \textbf{Best-fit} & \textbf{Median} & \textbf{84\%} & \textbf{16\%} \\
\noalign{\smallskip}
\tableline\tableline
\noalign{\smallskip}
\multicolumn{5}{c}{Model Parameters} \\
\noalign{\smallskip}
\hline
\noalign{\smallskip}
$\gamma_{\rm{HIRES}}$ & 4.8 & 5.4 & $+$1.6 & $-$1.6 \\
$\gamma_{\rm{SONG}}$ & 1.1 & 0.2 & $+$1.5 & $-$1.5 \\
$\gamma_{\rm{FEROS}}$ & -15.4 & -15.7 & $+$1.2 & $-$1.2 \\
$\gamma_{\rm{CORALIE}}$ & -5.4 & -5.0 & $+$1.2 & $-$1.2 \\
$\gamma_{\rm{HARPS}}$ & 8.1 & 8.8 & $+$1.5 & $-$1.5 \\
$\sigma_{\rm{HIRES}}$ & 2.71 & 2.68 & $+$0.85 & $-$0.80 \\
$\sigma_{\rm{SONG}}$ & 2.06 & 2.11 & $+$0.91 & $-$0.89 \\
$\sigma_{\rm{FEROS}}$ & 3.49 & 3.47 & $+$0.75 & $-$0.71 \\
$\sigma_{\rm{CORALIE}}$ & 1.88 & 2.50 & $+$0.75 & $-$0.64 \\
$\sigma_{\rm{HARPS}}$ & 2.41 & 2.69 & $+$0.75 & $-$0.63 \\
$z$ (ppm)  & 199.4 & 199.1 & $+$10.6 & $-$10.7 \\
$P$ (\rm{days}) & 14.2762 & 14.2767 & $+$0.0037 & $-$0.0037 \\
$T_0$ \rm{(BTJD)} & 1357.0135 & 1357.0149 & $+$0.0025 & $-$0.0026 \\
$b$ & 0.744 & 0.728 & $+$0.040 & $-$0.049 \\
$R_{p}/R_\star$ & 0.02846 & 0.02854 & $+$0.00084 & $-$0.00071 \\
$e \cos \omega$ & -0.054 & -0.028 & $+$0.063 & $-$0.061 \\
$e \sin \omega$ & -0.099 & -0.096 & $+$0.029 & $-$0.030 \\
K \rm{(m/s)} & 14.6 & 14.1 & $+$1.2 & $-$1.2 \\
$\rho_{\star} (\rhosun)$ & 0.06674 & 0.06702 & $+$0.00052 & $-$0.00052 \\
u1 & 0.12 & 0.35 & $+$0.36 & $-$0.24 \\
u2 & 0.71 & 0.44 & $+$0.30 & $-$0.44 \\
\noalign{\smallskip}
\hline
\noalign{\smallskip}
\multicolumn{5}{c}{Derived Properties} \\
\noalign{\smallskip}
\hline
\noalign{\smallskip}
$e$ & 0.113 & 0.115 & $+$0.034 & $-$0.030 \\
$\omega$ & -118.7 & -106.0 & $+$34.7 & $-$31.1 \\
$a$ \rm{(AU)} & 0.1233 & 0.1228 & $+$0.0025 & $-$0.0026 \\
$a/R_\star$ & 9.00 & 8.97 & $+$0.27 & $-$0.27 \\
$i$ ($^{\text{o}}$) & 85.67 & 85.75 & $+$0.36 & $-$0.35 \\
$\radp (\re)$ & 9.16 & 9.17 & $+$0.34 & $-$0.31 \\
$\radp (\rj)$ & 0.835 & 0.836 & $+$0.031 & $-$0.028 \\
$\massp (\me)$ & 63.4 & 60.5 & $+$5.7 & $-$5.7 \\
$\massp (\mj)$ & 0.200 & 0.190 & $+$0.018 & $-$0.018 \\
$\rhop$ \rm{(gcc)} & 0.455 & 0.431 & $+$0.064 & $-$0.060 \\
\noalign{\smallskip}
\tableline
\noalign{\smallskip}
\end{tabular}
\label{tab:planet}
\flushleft Notes: Parameters denote velocity zeropoints $\gamma$, radial velocity jitter $\sigma$, photometric zero point $z$, orbital period $P$, time of transit $T_{0}$, impact parameter $b$, star-to-planet radius ratio $R_{p}/R_\star$, eccentricity $e$, argument of periastron $\omega$, radial velocity semi-amplitude $K$, mean stellar density $\rho_{\star}$, linear and quadratic limb darkening coefficients u$_{1}$ and u$_{2}$, semi-major axis $a$, orbital inclination $i$, as well as planet radius ($\radp$), mass ($\massp$) and density ($\rhop$).
\end{table}

Figures \ref{fig:rvs} and \ref{fig:planet} show the radial velocity timeseries, phase-folded transit and RV data, and the corresponding best-fitting model. Table \ref{tab:planet} lists the summary statistics for all planet and model parameters. The system is well described by a planet in a 14.3 day orbit, which is nearly equal in size but $\sim$\,35\% less massive than Saturn ($\radp=\radplanet \rj$, $\massp=\massplanet \mj$), with tentative evidence for a mild eccentricity ($e=0.11\pm0.03$). The long transit duration ($\sim$\,0.5 days) is consistent with a non-grazing ($b\approx 0.7$) transit given the asteroseismic mean stellar density, providing further confirmation for a gas-giant planet orbiting an evolved star. The radial velocity data do not show evidence for any other short-period companions. Continued monitoring past the $\sim$4 orbital periods covered here will further reveal details about the orbital architecture of this system.

\section{Discussion}

\begin{figure*}
\begin{center}
\resizebox{\hsize}{!}{\includegraphics{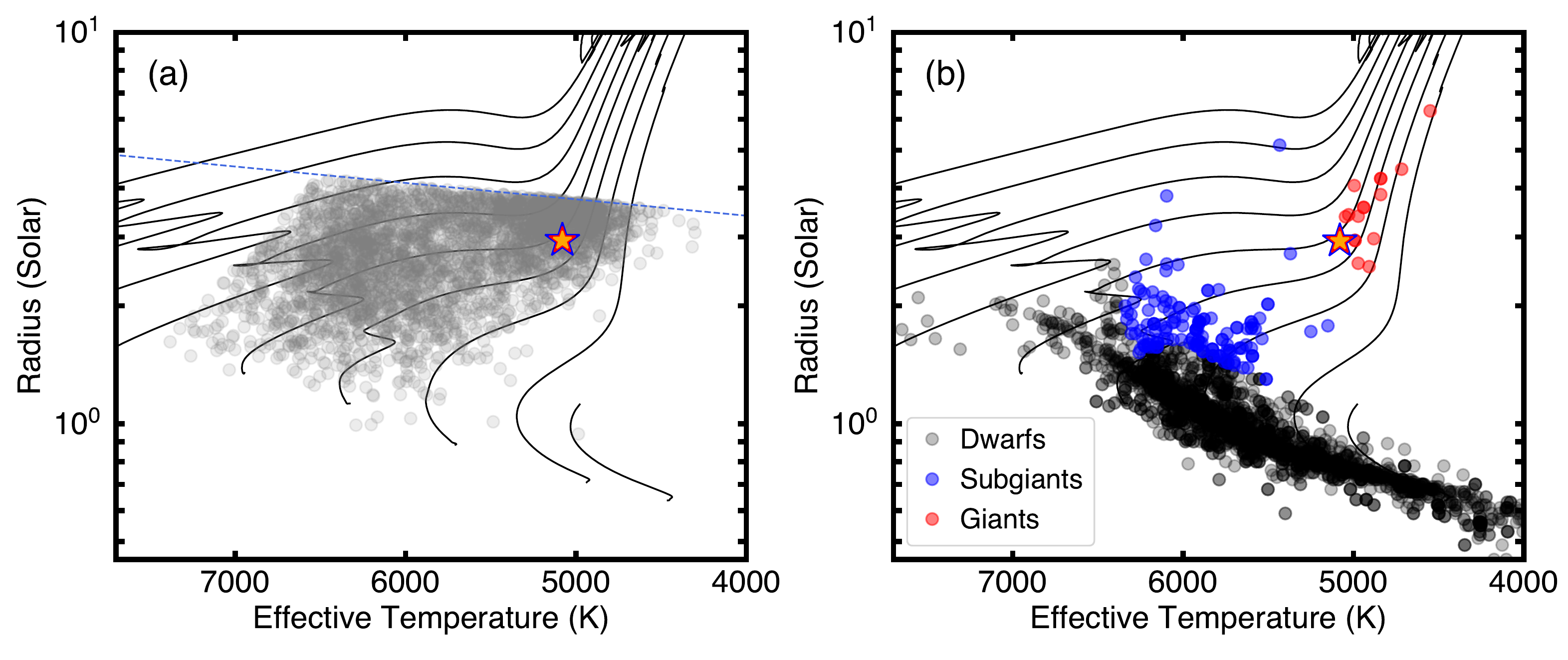}}
\caption{Stellar radius versus effective temperature for the expected \tess\ Cycle 1 yield of solar-like oscillators \citep[panel a,][]{schofield18} and for all stars with confirmed transiting planets (panel b). The blue dashed line in panel a marks the approximate limit below which 2-minute cadence data is required to sample the oscillations. Symbols in panel b are color-coded according to the evolutionary state of the star using solar-metallicity PARSEC evolutionary tracks. \target\ falls on the border between subgiants and red-giants, and is highlighted with an orange/red/blue star symbol. \target\ is a typical target for which we expect to detect solar-like oscillations with \tess, but occupies a rare parameter space for an exoplanet host.}
\label{fig:hrd}
\end{center}
\end{figure*}

\planet\ joins an enigmatic but growing class of transiting planets orbiting stars that have significantly evolved off the main sequence. Figure \ref{fig:hrd} compares the position of \target\ within the expected population of solar-like oscillators to be detected with \tess\ (panel a) and within the known population of exoplanet host stars. Evolutionary states in Figure \ref{fig:hrd}b have been assigned using solar-metallicity PARSEC evolutionary tracks \citep{bressan12} as described in \citet{berger18}$^{6}$\footnote{$^{6}$see also \url{https://github.com/danxhuber/evolstate}}. 
\target\ sits at the boundary between subgiants and red giants, with the measured \dnu\ value indicating that the star has just started its ascent on the red-giant branch \citep{mosser14}.
\target\ is a typical target for which we expect to detect solar-like oscillations with \tess, predominantly due to the increased oscillation amplitude, which are well known to scale with luminosity \citep{kb95}. On the contrary, \target\ is rare among known exoplanet hosts: while radial velocity searches have uncovered a large number of planets orbiting red giants on long orbital periods \citep[e.g.][]{wittenmyer11}, less than 15 transiting planets are known around red-giant stars (as defined in Figure \ref{fig:hrd}b). \target\ is the sixth example of a transiting planet orbiting a late subgiant / early red giant with detected oscillations, following Kepler-91 \citep{barclay12b}, Kepler-56 \citep{steffen12c,huber13b}, Kepler-432 \citep{quinn15}, K2-97 \citep{grunblatt16} and K2-132 \citep{grunblatt17,jones18}. 

\begin{figure*}
\begin{center}
\resizebox{\hsize}{!}{\includegraphics{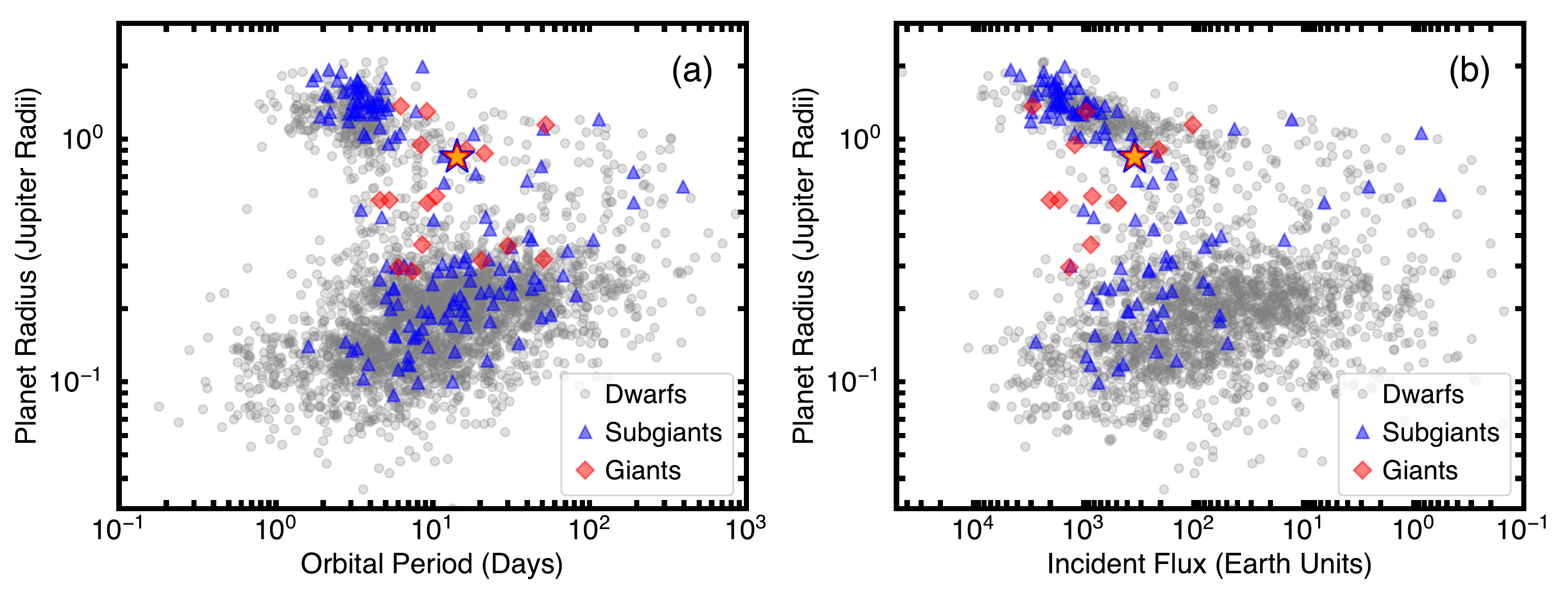}}
\caption{Planet radius versus orbital period (panel a) and incident flux (panel b) for confirmed exoplanets. Symbols are color-coded according to the evolutionary state of the host star (see Figure \ref{fig:hrd}). \target\ b is highlighted in both panels with an orange/red/blue star symbol.}
\label{fig:incflux}
\end{center}
\end{figure*}

Transiting planets orbiting evolved stars are excellent systems to advance our understanding of the effects of stellar evolution on the structure and evolution of planets \citep[see e.g.][for a review]{veras16}. For example, such systems provide the possibility to test the effects of stellar mass, evolution and binarity on planet occurrence \citep[e.g.][]{johnson10,schlaufman13,stephan18}, which are still poorly understood. Furthermore, the increased irradiance on the planet caused by the evolution of the host star has been proposed as a means to distinguish between proposed mechanisms to explain the inflation of gas-giant planets beyond the limits expected from gravitational contraction and cooling \citep{hubbard02,lopez15}. Recent discoveries by the K2 mission have indeed yielded evidence that planets orbiting low-luminosity RGB stars are consistent with being inflated by the evolution of the host star \citep{grunblatt16,grunblatt17}, favoring scenarios in which the energy from the star is deposited into the deep planetary interior \citep{bodenheimer01}. 

Based on its radius and orbital period, \target\ would nominally be classified as a warm Saturn, sitting between the well-known population of hot Jupiters and the ubiquitous population of sub-Neptunes uncovered by Kepler (Figure \ref{fig:incflux}a). Taking into account the evolutionary state of the host star, however, \target\ falls at the beginning of the ``inflation sequence'' in the radius-incident flux diagram (Figure \ref{fig:incflux}b), with planet radius strongly increasing with stellar incident flux \citep{kovacs10,demory11,miller11,thorngren18}. Since \planet\ is currently not anomalously large compared to the observed trend and scatter for similar planets (Figure \ref{fig:incflux}b) and low-mass planets are expected to be more susceptible to planet reinflation \citep{lopez15}, \target\ may be a progenitor of a class of re-inflated gas-giant planets orbiting RGB stars. 

If confirmed, the mild eccentricity of \planet\ would be consistent with predictions of a population of planets around evolved stars for which orbital decay occurs faster than tidal circularization \citep{villaver14,grunblatt18}. Moreover, combining the asteroseismic age of the system with the possible non-zero eccentricity would allow constraints on the tidal dissipation in the planet, which drives the circularization of the orbit. Using the formalism by \citet{mardling11} \citep[see also][]{gizon13,davies15,ceillier16}, the current constraints would imply a minimum value of the planetary tidal quality factor $Q_{p;{\rm min}}\approx 3.2 \times 10^{4}$, below which the system would have been already circularized in $\sim$\,5\,Gyr. Compared to the value measured in Saturn \citep[$Q\approx1800$,][]{lainey17}, this would demonstrate the broad diversity of dissipation observed in giant planets. Since tidal dissipation mechanisms vary strongly with internal structure \citep[see e.g.][]{guenel14,ogilvie14,andre17}, this may also contribute to understanding the internal composition of such planets. We caution, however, that further RV measurements will be needed to confirm a possible non-zero eccentricity for \planet.

\begin{figure}
\begin{center}
\resizebox{\hsize}{!}{\includegraphics{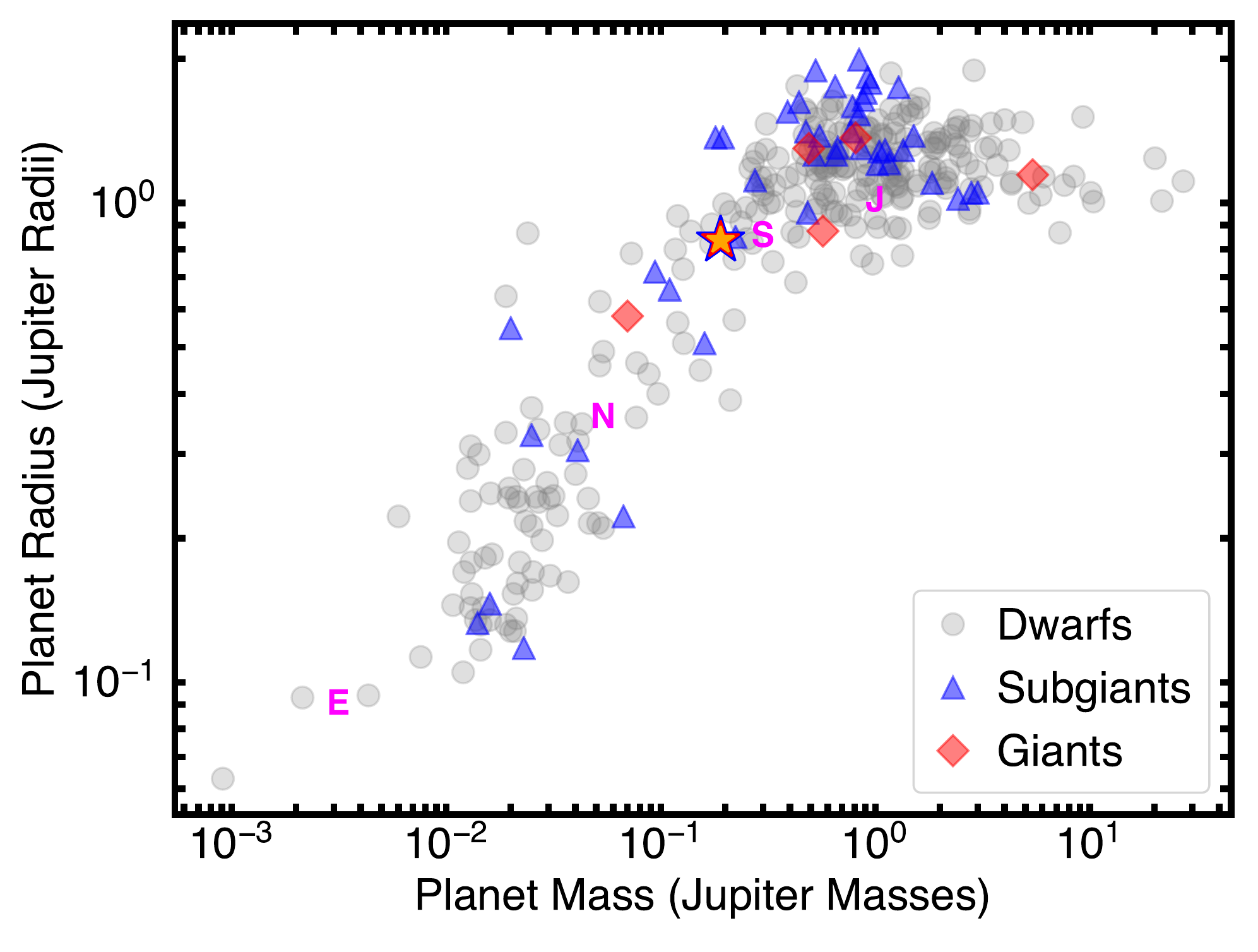}}
\caption{Mass-radius diagram for confirmed planets with densities measured to better than 50\%. Symbols are color-coded according to the evolutionary state of the host star (see Figure \ref{fig:hrd}). \target\ b is highlighted with a orange/red/blue star symbol. Magenta letters show the position of solar system planets.}
\label{fig:massrad}
\end{center}
\end{figure}

The precise characterization of planets orbiting evolved, oscillating stars also provides valuable insights into the diversity of compositions of planets through their mean densities. \planet\ falls in the transition region between Neptune and sub-Saturn sized planets for which radii increase as $R_{P} \approx M_{P}^{0.6}$, and Jovian planets for which radius is nearly constant with mass \citep[][Figure \ref{fig:massrad}]{weiss13,chen17}. Recent studies of a population of sub-Saturns in the range $\sim$\,4--8\,\re\ also found a wide variety of masses, approximately 6--60\,\me, regardless of size \citep{petigura17, vaneylen18b}. Furthermore,  masses of sub-Saturns correlate strongly with host star metallicity, suggesting that metal-rich disks form more massive planet cores. 
\planet\ demonstrates that this trend does not appear to extend to planets with sizes $>8$\,\re, given its mass of $\sim$\,60\,\me\ and a roughly sub-solar metallicity host star ($\feh \approx -0.08$\,dex).   
This suggests that Saturn-sized planets may follow a relatively narrow range of densities, a possible signature of the transition in the interior structure \citep[such as the increased importance of electron degeneracy pressure,][]{zapolsky69} leading to different mass-radius relations between sub-Saturns and Jupiters. 
We note that \planet\ is one of the most precisely characterized Saturn-sized planets to date, with a density uncertainty of $\sim$\,15\%. 

\section{Conclusions}

We have presented the discovery of \planet, the first transiting planet orbiting an oscillating host star identified by \tess.
Our main conclusions are as follows:

\begin{itemize}
    \item \target\ is a late subgiant / early red giant with a clear presence of mixed modes. Combined spectroscopy and asteroseismic modeling revealed that the star has just started its ascent on the red giant branch, with $\rstar = \radstar \rsun$, $\mstar = \massstar  \msun$ and near-solar age (\agestar\,Gyr). \target\ is a typical oscillating star expected to be detected with \tess, and demonstrates the power of asteroseismology even with only 27 days of data.
    
    \item The oscillation amplitude of \target\ is consistent with ensemble measurements from \kep. This confirms that the redder bandpass of \tess\ compared to \kep\ only has a small effect on the oscillation amplitude \citep[as expected from scaling relations,][]{kb95,ballot11}, supporting the expected yield of thousands of solar-like oscillators with 2-minute cadence observations in the nominal \tess\ mission \citep{schofield18}. A detailed study of the asteroseismic performance of \tess\ will have to await ensemble measurements of noise levels and amplitudes. 
    
    \item \planet\ is a ``hot Saturn'' ($F=\incplanet \fe$, $\radp=\radplanet \rj$, $\massp=\massplanet \mj$) and joins a small but growing population of close-in, transiting planets orbiting evolved stars. Based on its incident flux, radius and mass, \planet\ may be a precursor to the population of gas giants that undergo radius re-inflation due to the increased irradiance as their host star evolves up the red-giant branch.
    
    \item \planet\ is one the most precisely characterized Saturn-sized planets to date, with a density measured to $\sim$\,15\%. \planet\ does not follow the trend of increasing planet mass with host star metallicity discovered in sub-Saturns with sizes between $4-8\,\re$, which has been linked to metal-rich disks preferentially forming more massive planet cores \citep{petigura17}. The moderate density ($\rhop = \denplanet$\,g\,cm$^{-3}$) suggests that Saturn-sized planets may follow a relatively narrow range of densities, a possible signature of the transition in the interior structure leading to different mass-radius relations for sub-Saturns and Jupiters.    
   
\end{itemize}

\target\ provides a first glimpse at the strong potential of \tess\ to characterize exoplanets using asteroseismology. \planet\ has one the most precisely characterized densities of known Saturn-sized planets to date, with an uncertainty of $\sim$\,15\%. Thanks to asteroseismology the planet density uncertainty is dominated by measurements of the transit depth and the radial velocity amplitude, and thus can be expected to further decrease with continued transit observations and radial velocity follow-up, which is readily performed given the brightness (V=8) of the star. Ensemble studies of such precisely characterized planets orbiting oscillating subgiants can be expected to yield significant new insights on the effects of stellar evolution on exoplanets, complementing current intensive efforts to characterize planets orbiting dwarfs.

\acknowledgments
The authors wish to recognize and acknowledge the very significant cultural role and reverence that the summit of Maunakea has always had within the indigenous Hawai`ian community.  We are most fortunate to have the opportunity to conduct observations from this mountain.
We thank Andrei Tokovinin for helpful information on the Speckle observations obtained with SOAR.
D.H.\ acknowledges support by the National Aeronautics and Space Administration through the TESS Guest Investigator Program (80NSSC18K1585) and by the National Science Foundation (AST-1717000). A.C.\ acknowledges support by the National Science Foundation under the Graduate Research Fellowship Program.
W.J.C., W.H.B., A.M., O.J.H.\ and G.R.D.\ acknowledge support from the Science and Technology Facilities Council and UK Space Agency.
H.K.\ and F.G.\ acknowledge support from the European Social Fund  via the Lithuanian Science Council grant No.\ 09.3.3-LMT-K-712-01-0103. 
Funding for the Stellar Astrophysics Centre is provided by The Danish National Research Foundation (Grant DNRF106).
A.J.\ acknowledges support from FONDECYT project 1171208, CONICYT project BASAL AFB-170002, and by the Ministry for the Economy, Development, and Tourism's Programa Iniciativa Cient\'{i}fica Milenio through grant IC\,120009, awarded to the Millennium Institute of Astrophysics (MAS). R.B.\ acknowledges support from FONDECYT Post-doctoral Fellowship Project 3180246, and from the Millennium Institute of Astrophysics (MAS).
A.M.S.\ is supported by grants ESP2017-82674-R (MINECO) and SGR2017-1131 (AGAUR).
R.A.G.\ and L.B.\ acknowledge the support of the PLATO grant from the CNES. The research leading to the presented results has received funding from the European Research Council under the European Community's Seventh Framework Programme (FP7\/2007-2013) \/ ERC grant agreement no 338251 (StellarAges).
S.M.\ acknowledges support from the European Research Council through the SPIRE grant 647383.
This work was also supported by FCT (Portugal) through national funds and by FEDER through COMPETE2020 by these grants: UID/FIS/04434/2013 \& POCI-01-0145-FEDER-007672, PTDC/FIS-AST/30389/2017 \& POCI-01-0145-FEDER-030389.
T.L.C.\ acknowledges support from the European Union's Horizon 2020 research and innovation programme under the Marie Sk\l{}odowska-Curie grant agreement No.~792848 (PULSATION).
E.C.\ is funded by the European Union’s Horizon 2020 research
and innovation program under the Marie Sklodowska-Curie
grant agreement No. 664931.
V.S.A.\ acknowledges support from the Independent Research Fund Denmark (Research grant 7027-00096B).
D.S.\ acknowledges support from the Australian Research Council.
S.B.\ acknowledges NASA grant NNX16AI09G and NSF grant AST-1514676.
T.R.W.\ acknowledges support from the Australian Research Council through grant DP150100250.
A.M.\ acknowledges support from the ERC Consolidator Grant funding scheme (project ASTEROCHRONOMETRY, G.A. n. 772293).
S.M.\ acknowledges support from the Ramon y Cajal fellowship number RYC-2015-17697.
M.S.L.\ is supported by the Carlsberg Foundation (Grant agreement no.: CF17-0760).
A.M.\ and P.R.\ acknowledge support from the HBCSE-NIUS programme.
J.K.T.\ and J.T.\ acknowledge that support for this work was provided by NASA through Hubble Fellowship grants HST-HF2-51399.001 and HST-HF2-51424.001 awarded by the Space Telescope Science Institute, which is operated by the Association of Universities for Research in Astronomy, Inc., for NASA, under contract NAS5-26555.
T.S.R.\ acknowledges financial support from Premiale 2015 MITiC (PI B. Garilli).
This project has been supported by the NKFIH K-115709 grant and the Lend\"ulet Program of the Hungarian Academy of Sciences, project No. LP2018-7/2018.

Based on observations made with the Hertzsprung SONG telescope operated on the Spanish Observatorio del Teide on the island of Tenerife by the Aarhus and Copenhagen Universities and by the Instituto de Astrofísica de Canarias.
Funding for the TESS mission is provided by NASA's Science Mission directorate. We acknowledge the use of public TESS Alert data from pipelines at the TESS Science Office and at the TESS Science Processing Operations Center. This research has made use of the Exoplanet Follow-up Observation Program website, which is operated by the California Institute of Technology, under contract with the National Aeronautics and Space Administration under the Exoplanet Exploration Program. This paper includes data collected by the TESS mission, which are publicly available from the Mikulski Archive for Space Telescopes (MAST).

\software{Astropy \citep{astropy}, Matplotlib \citep{matplotlib}, DIAMONDS \citep{corsaro14}, isoclassify \citep{huber17}, EXOFASTv2 \citep{eastman17}, ktransit \citep{ktransit}}

\bibliography{references.bib}

\end{document}